# Skyrmion-like states in multilayer exchange coupled ferromagnetic nanostructures with distinct anisotropy directions

A.A. Fraerman[1,2], O.L.Ermolaeva[1,2], E.V.Skorohodov[1], N.S.Gusev[1], V.L.Mironov[1,2], S.N.Vdovichev[1,2], E.S.Demidov[2]

[1] Institute for physics of microstructures RAS, Nizhny Novgorod, GSP-105, 603950, Russia
[2] Lobachevsky State University of Nizhny Novgorod, Nizhny Novgorod, Gagarin Avenue, 23, 603950, Russia

**Abstract** We report the experimental observation of magnetic skyrmion-like states in patterned ferromagnetic nanostructures consisting of perpendicular magnetized Co/Pt multilayer film exchange coupled with Co nanodisks in vortex state. The magnetic force microscopy and micromagnetic simulations show that depending on the magnitude of Co/Pt perpendicular anisotropy in these systems two different modes of skyrmion formation are realized.

Magnetic skyrmion is a localized spin configuration demonstrating unusual topological and transport properties [1]. This state was predicted theoretically as the effect of Dzjaloshinskii-Moriya interaction in crystals without an inversion center [2-4]. Experimentally the skyrmion lattices were observed in some crystals (MnSi, FeGe and other) at low temperatures [5-7]. Now one of actual problem is to expand the class of magnetic materials suitable for realization of skyrmions stable at room temperature. Recently the formation of skyrmion-like states induced by magnetic vortex in artificial ferromagnetic nanostructures was considered theoretically in Ref. [8]. In current letter we present the experimental realization of skyrmion-like states in Co/Pt multilayer films exchange coupled with Co nanodisks (Co/Pt-Co$_{disk}$ nanostructures).

The initial thin film structures [Co (0.5 nm) / Pt (1 nm)]$_5$ (denoted further as Co/Pt) and [Co (0.5 nm) / Pt (1 nm)]$_5$ / Co (denoted further as Co/Pt-Co) was grown by DC magnetron sputtering on Si substrate with Ta (10 nm) / Pt (10 nm) buffer layer. The magnetic properties of Co/Pt and Co/Pt-Co thin film structures were investigated by magneto-optical Kerr effect (MOKE) and ferromagnetic resonance (FMR) methods. In experiments we used two Co/Pt structures differing by coercivity. First structure (structure I) had the coercive field $H_{cI} = 130$ Oe and saturation field $H_{sI} = 210$ Oe, while the second one (structure II) had coercive field $H_{cII} = 180$ Oe and saturation field $H_{sII} = 300$ Oe. The corresponding hysteresis curves are presented in Fig. 1a.

Covering Co/Pt structures by Co layer with thickness $t_{Co} < 1.5$ nm led to narrowing hysteresis loop and for $t_{Co} > 1.5$ nm we registered the anhysteretic magnetization curves. For example, normalized MOKE remagnetization loop for the structure I covered by 20 nm Co layer is presented in Fig. 1b. The FMR measurements showed that Co/Pt-Co thin film structures consist of two coupled effective oscillators Co/Pt (easy axis anisotropy) and Co (easy plane anisotropy) with surface energy of exchange interaction $J = 1.9 \times 10^{-3}$ J/m$^2$ and we believe that anhysteretic behavior of these structures with $t_{Co} > 1.5$ nm shows that easy plane anisotropy is dominant.

Removing Co coating layer by ion etching with ion energy 200 eV does not destroy Co/Pt multilayer structure. The hysteresis loop for Co/Pt multilayer film after 20 nm Co coating removal is presented in Fig. 1c.

The Co/Pt-Co$_{disk}$ nanostructures with array of circle nanodisks (disk diameter 200 nm, thickness 20 nm, period 400 nm) were fabricated by electron beam lithography and ion etching (ion energy 200 eV, ion beam current 5 mA) of structures I and II with 20 nm Co upper layer.

The magnetic states and the magnetization reversal effects in these nanostructures were studied using a vacuum multimode magnetic force microscope (MFM)

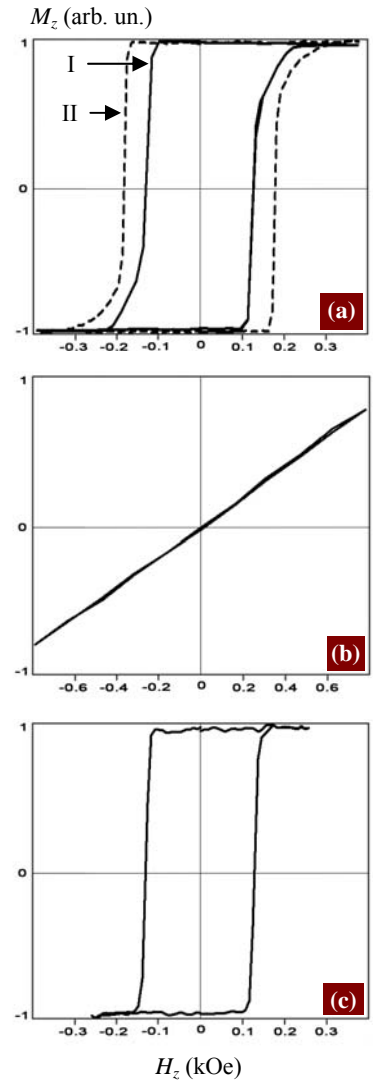

**Fig. 1.** (a) Normalized hysteresis curves of Co/Pt multilayer films with different coercivity (structures I and II). (b) Hysteresis loop of Co/Pt - Co (20 nm) thin film structure. (c) Hysteresis curve of Co/Pt multilayer film (structure I) after Co (20 nm) removal.

"Solver-HV", which was equipped with dc electromagnets. The MFM measurements were performed in two-pass mode with relief extraction.

The MFM probe was magnetized along the axis so that its magnetic moment was directed towards the sample. The phase shift of cantilever oscillations under the gradient of the magnetic force was registered to obtain the MFM contrast [9]. All measurements were performed in a vacuum of $10^{-4}$ Torr, which improved the MFM signal due to an increase in the cantilever quality factor.

The experiments were carried out in the following scenario. At the first stage the samples were magnetized in an external magnetic field $H_0 = 10$ kOe, so that the magnetization direction coincided with the direction of the outward normal to the sample surface. Then at the second stage the reversed magnetic field $H_R$ was applied with amplitude near $H_c$ of Co/Pt and remagnetization effect was registered.

We observed experimentally that in Co/Pt-Co$_{disk}$ systems two different modes of magnetization switching were implemented depending on the parameters of Co/Pt multilayer structure. For structure I the remagnetization effect was observed in the field 240 Oe greater than the saturation field ($H_R > H_{sI}$). The MFM images of the initial state and state after remagnetization for structure I are presented in Fig. 2a,b. A characteristic feature of MFM image presented in Fig. 2a is a narrow bright spot in the center of the disk. After magnetization switching the MFM image was changed dramatically and we observed the broad bright spot in the MFM contrast distribution indicating a change in magnetic structure of the sample (Fig. 2b).

On the other hand, the remagnetization of structure II has some difference. The switching effect was observed in the field 170 Oe lower than the coercive field ($H_R < H_{cII}$). Typical MFM images of structure II before and after remagnetization are shown in Fig. 3 a,b. In this case we observed the appearance of broad dark spot in the MFM contrast distribution over the disk (Fig. 3b).

To explain observed effects we supposed that remagnetization of structure I is connected with remagnetizing the peripheral region of the Co/Pt film (outside the Co disk), while remagnetization of structure II is caused by reorientation of magnetization in central part of Co disk and underlying Co/Pt area. It is known that in free Co disk with such dimensions the vortex state is realized [10]. So we assume that after initial magnetizing both Co disk and exchange coupled underplaying region of Co/Pt have vortex distribution of magnetization [8] (see Fig. 2c and Fig. 3c). In this case the MFM registered lateral scale of non-homogeneity for Z component of stray field is defined by total thickness of Co/Pt-Co$_{disk}$ structure, which is equal about 28 nm. The remagnetization in structure I is accompanied by reorientation of magnetization in Co/Pt out of Co disk. In this case magnetization in central and peripheral parts has opposite orientation and the characteristic scale of stray field localization is defined by disk diameter (about 200 nm). These changes of stray field localization lead to the observed effect of MFM contrast broadening. The remagnetization effect in structure II is defined by reorientation of central part of Co disk and underlying Co/Pt. In this case the distribution of magnetization (Fig. 3d) coincides with distribution for structure I (Fig. 2d) but has opposite orientation of magnetic moments. As a consequence we observe the opposite contrast in MFM images.

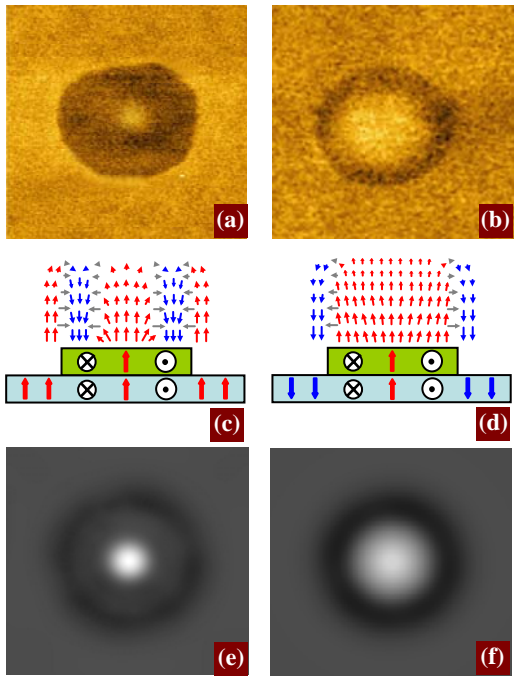

**Fig. 2.** (a) Normalized experimental MFM image of initial state (first stage) in structure I. (b) Experimental MFM image of final state (second stage) after remagnetization. (c), (d) Schematic drawings of magnetic state and stray fields corresponding to the first and second stages respectively. (e), (f) Model MFM images of initial and final states respectively. The frame size 350 × 350 nm.

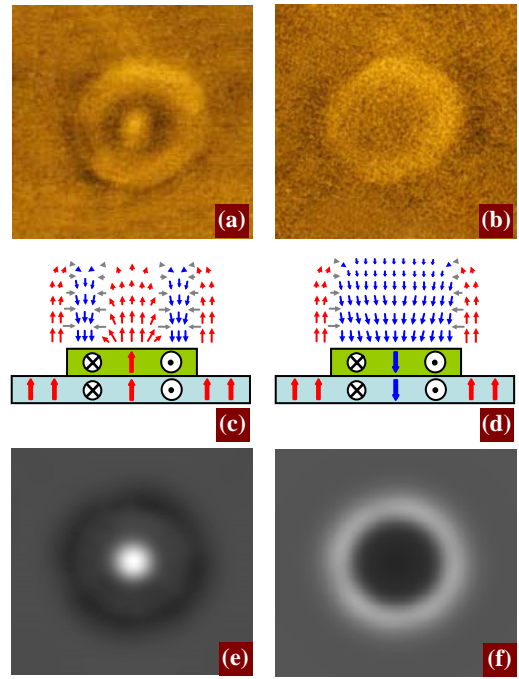

**Fig. 3.** (a) Normalized experimental MFM image of initial state (first stage) in structure II. (b) Experimental MFM image of final state (second stage) after remagnetization. (c),(d) Schematic drawings of magnetic state and stray fields corresponding to the first and second stages respectively. (e) (f) Model MFM images of initial and final states respectively. The frame size 350 × 350 nm.

To verify the proposed scenario of remagnetization we performed additional micromagnetic modeling using a standard object oriented micromagnetic framework code (OOMMF) [11] based on numerical solution of the Landau-Lifshitz-Gilbert equation. We considered the system of Co/Pt-Co$_{disk}$ with following parameters: the Co/Pt part was $400 \times 400 \times 7.5$ nm, Co disk was $200 \times 20$ nm. The calculations were carried out for the following Co parameters: the exchange stiffness was $A_{Co} = 2.5 \times 10^{-11}$ J/m, the saturation magnetization was $M_{Co} = 14 \times 10^5$ A/m, anisotropy constant was equal to zero $K_{Co} = 0$. The Co/Pt multilayer structure was modeled as a continuous anisotropic medium with the following parameters: the exchange stiffness was $A_{Co/Pt} = 1.5 \times 10^{-11}$ J/m, the saturation magnetization was $M_{Co/Pt} = 5 \times 10^5$ A/m, anisotropy constant was $K_{Co/Pt} = 4 \times 10^5$ J/m$^3$ (for structure I) and $K_{Co/Pt} = 5 \times 10^5$ J/m$^3$ (for structure II). Parameters of the system I coincide with the parameters used in [8], but for the system II the anisotropy constant was taken a bit more than for the system I. The exchange stiffness of Co/Pt – Co interlayer interaction was taken as $A_{Co/Pt-Co} = 2 \times 10^{-11}$ J/m.

As it was shown in micromagnetic simulations Co disk and exchange coupled region of Co/Pt film lying directly under the disk have the vortex-like distribution of magnetization. Interlayer interaction of two materials with distinct anisotropy directions leads to the appearance of significant Z component in vortex shell magnetization both for Co disk and for Co/Pt underlying area. As a consequence the model MFM image of this state has the narrow bright spot in the center of the disk (Fig. 2e and Fig. 3e), which agrees qualitatively with the experimental pictures (Fig. 2a and Fig. 3a).

In micromagnetic modeling of structure I the magnetization switching was observed after applying 350 Oe reversed field. The central region of the vortex distribution in Co disk and underplaying Co/Pt retained the direction of magnetization, while the peripheral region of the Co/Pt film (outside the disk) taken the opposite magnetization. So, the formation of skyrmion-like state in the Co/Pt layer, similar to that described in [8] was observed. The structure II demonstrated some different micromagnetic behavior. Increasing the perpendicular anisotropy of the Co/Pt layer led to the stabilization of the peripheral area and the magnetic state switching was observed at 280 Oe by changing magnetization direction in the vortex core region. In this case the skyrmion-like state but with opposite orientation (in comparison with structure I) is formed. As a result the corresponding MFM image has the opposite contrast in the central region of Co disk (Fig. 3f), which agrees qualitatively with the experimental data.

Thus, magnetic force microscopy and micromagnetic simulations show that in patterned systems Co/Pt-Co$_{disk}$ the skyrmion-like states are realized. Depending on the Co/Pt coercivity, associated with the magnitude of perpendicular anisotropy, two different modes of skyrmion formation were demonstrated.

Such structures with well-controlled skyrmion-like distribution of magnetization are very promising for fundamental studies of spin-dependent transport peculiarities and magnetodynamic phenomena specific to strongly inhomogeneous magnetic systems.

The authors are very thankful to A.Yu. Klimov, V.V. Rogov and A.E. Pestov for assistance. This work was supported by Russian Foundation for Basic Research (projects 11-02-00434 and 11-02-00589) and Ministry of Education and Science RF (agreement 02.B.49.21.0003).